# Media Education as Theoretical and Practical Paradigm for Digital Literacy: An Interdisciplinary Analysis

José Gómez Galán[a]

[a] *Director of CICIDE (Metropolitan University, AGMUS, Puerto Rico-United States & Catholic University of Avila, Spain).
Corresponding author: jogomez@suagm.edu / jose.gomez@ucavila.es*

**Abstract**

In this article we offer an analysis of the practical and theoretical paradigm of media education as a fundamental pedagogical model for the adequate development of the current methods of digital literacy. In a society dominated by the flow of information and communicative processes -in which digitalization has led to techno-media convergence, the complete uniting of ICT novelties and traditional media- it is essential to use all the vast experience of basic principles, pedagogical theories and practice which this educational paradigm has offered during decades.

The current techno-media society is not an altruistic system; the communicative processes continue to be dominated by economic, political and social elites whose main interest is to influence and control the population. In this context -conceiving education as the only way to achieve the full and democratic development of our society, for its growth in values and solidarity- an analysis of reality starting in school is essential by forming a citizenship conscious of the power and influence of the ICT and its true meaning in the world. An authentic digital literacy should include the correct understanding of the new techno-media languages, and cannot simply be reduced to a formation of a technical and instrumental kind[*].

## 1. Introduction and Question Statement: Searching for Efficient Educational Strategies for Digital Literacy in a Techno-Media Society

We are finding ourselves in a historically amazing moment. The digital revolution is profoundly modifying our lifestyle habits, our means of understanding the world. It seems evident that we are aiding the birth of a new stage in Western civilization. Through the neolithic and industrial revolutions, the digital revolution of ICT –this has brought about the digitalization of all information–, is setting the stage for the third great stage: the society of communication. Physical space, as much a determining factor as in the previous historic stages, stops having such relevance when it is substituted by a virtual space where communications between human beings cover the whole planet, globalizing it in every scope (cultural, economic, social, etc.). The flows of information constitute extremely complicated communicative networks that penetrate everything and decidedly impact our way of living, already so differently than in the past. And everything is just the start of a future that we can glimpse in the distance. In a few decades, more changes will have been produced than in previous centuries, and the evolution of this digital world continues in frenetic progression.

In this new context, education has to offer responses to new needs. The human being of the 21st century needs knowledge, capabilities, and outlooks for a new society in which the communicative processes have so much relevance. It is for this reason that we are currently speaking insistently of the need for digital literacy. Just like traditionally, when one of the basic objectives of education was linguistic and cognitive literacy – with the acquisition of fundamental competencies such as, for example, the ability to read and to write – today, literacy in the new multimedia and hypermedia languages, born in the digital revolution, are also

---

necessary. And the necessary capabilities for using the new tools that make their use and development possible. What we would definitively call digital literacy.

Currently there is abundant scientific literature which attempts to propose answers to what could be the main directives for carrying out adequate digital literacy. There are some important recent contributions according to the educational level of interest, such as Arrow and Finch [1], centered on initial multimedia literacy, for children, in the school environment as well as in the family one; Alvarez, Salavati, Nussbaum and Milrad [2] in primary education offering the results of pedagogical experience in the classroom; Akkoyunlu and Yilmaz [3], focused on the efficiency of teaching staff in this area; Bulfin and Koutsogiannis [4], in secondary education, for adolescents; Goodfellow [5], Olaniyan, Graham and Nielsen [6], Lea [7], centered on higher education. Also to be mentioned is the work of Littlejohn, Beetham and McGill [8] and of Gainer [9], for establishing the state of the question and new approaches to digital literacy.

We consider, however, that on too many occasions the trees don't stop seeing the forest, and in too many research figures from the international scientific community the problem itself certainly appears limited, if only from an instrumental and technical perspective. Digital literacy is reduced to try to develop didactic strategies for the acquisition of competencies in the management of computers, digital whiteboards, tablets, smartphones, operating systems, information searches on the Internet, etc., that is, basic training in hardware and software from a primarily operational dimension.

But the problem today is much greater and should be confronted from its roots. The majority of the school-age population continually acquires digital capabilities in their daily lives, since ICT dominates everything. Except in the cases of digital literacy in the adult population, the didactic instrumental focus is not necessary. The children and young adults of the communication society are perfectly familiar with the employment of new media. The problem, without a doubt, is something else: that these new media decisively condition their lives, they influence them, they affect them, they guide their habits, their interests, their opinions, etc. They are definitely omnipresent in their world; they are their world. And what results is important to highlight: in some way, they are the "new media." As such, the educational strategies that we need should be consistent with this and face the real problem.

We will explain in detail the reason for all of this. We start from the fact that we have developed widely in other contributions (Gómez Galán, [10] [11]; Gómez Galán y Mateos [12]), that at present we are, due to the digital revolution created by information and communication technology (ICT), in a process of convergence that we call *techno-media*, and in which media –as much the traditional as the newest, from the press or the radio to the Internet and social networks- stop existing as separate entities in order to form part of a unique digital media that covers the whole of human communication and in which, without a doubt, we find ourselves at its dawn. In the future, there will be only one medium –with many manifestations, naturally, but participating as one essence– in which we will not be able to speak of distinct communicative and/or technological confines. Watching television will mean interacting with what today we do with social networks, reading the press at the same time that we create our own content, listening to the radio while we envision the participation of talk show guests, and we will communicate with each other through videoconferencing with them, etc., the divisions between media will not exist and we will find ourselves in a unique interactive communication system, born of the digital paradigm.

It will even create a convergence of hardware and software, and we will not have a variety of them available to us for the transmission of the different media-technologies, the traditional ones, and we'll call them the new ones. The medium-system will be the same, the only thing that will change will be the size of this medium that we will have at our disposition, through the Internet/Digital paradigm, and we will have all the existing media. The only difference will be the comfort of viewing, portability, etc. But they will have the same functions. That is, we will not talk about different media, but rather of one medium with different sizes adapted to our needs.

But in this new reality, let's not trick ourselves, the structure of power –economic, social, political, etc.– of the classic communication media will not disappear. They will just have been adapted to a new technological framework, but they will continue exercising their influence and domination over citizens, even more intensely than currently, since they will be offering an apparent -yes, apparent- freedom to the user-consumer-voter, who, thanks to the development of digital technologies, will believe that they have widely gained independence and decision-making. But this is just an illusion created by means of super-sophisticated communication channels that are controlled, without a doubt, by the same powers that be as always. For this reason, more than for a society of communication –in that the term might denote complete freedom in the flows of communication– it would be necessary to talk of a techno-media society. A society in which the digital revolution, in some way as a result of an adaptation, in the continual processes that The

History of Humanity supposes, of the power structures to new scenery in which physical space stops being decisive. And which, as always, supposes enormous benefits in all fields: economic, social, power.

The educational problem of forming our future citizens continues, therefore, including the new situation that emerged in the 20th century at the start of the emergence of mass communication media. The digital revolution has promoted its effects enormously, since it has brought the power of communication and its impact on society to extraordinary limits. But this is not a completely new problem. It's the same problem, but increased. And what needs to change is the authentic challenge of digital literacy.

## 2. Digital Literacy after Media Education

What would be, therefore, the necessary educational and pedagogical strategies for this? What would be the main guidelines that we need to develop today within the scope of *digital literacy*? Undoubtedly, there are those that give us a global response to the main problem: the education of new generations, not only for their integration into society, like what is habitually defended, but for the betterment of it. And this is not achieved, in any way, by training only the children and young adults in order to widen and optimize their instrumental abilities. It is reached by being critical of everything that impedes the global development of society, which generates injustices, inequalities, which does not face the problems affronting humanity and the whole of our planet. But what is necessary is knowledge of our world without interferences and interests created to exclusively benefit the elites.

In this sense, the knowledge of technology and means of communication, the absolute protagonists of our world –in a global way and in all cases previously described-- becomes essential. And the most important is in the world of education, to face the challenge, it is not necessary to create an extraordinary and complex innovation of uncertain productivity, since we already have a fundamental pedagogical paradigm of long-standing tradition and excellent results where it is applied: media education. This would answer every problem that we have presented and would serve, without a doubt, as a base of an authentic digital literacy education that would respond to the true needs of society.

*Media education* is a scientific discipline within the realm of education sciences, which practically came about with the origin of the mass communication media. It is the most widespread term, although we personally prefer to use the term *media pedagogy*. Recently, and above all in the United States, the term *media literacy* (Potter [13]; Kellner & Share [14]) has gained ground, although in any case –except in very minor aspects—we are speaking about the same educational paradigm. It is intended primarily for the education of citizens for the adequate use and consumption of techno-media products and also so that they can reach the capacities to analyze, use, and even express, in different ways, the message produced by them. Media education proposes as much a theoretical dimension –the study of the problems involved in dialogue with the rest of the educational sciences-- as a practical dimension-basically to establish the most accurate didactic strategies and the integration of them in educational contexts-. It should not be confused with education through media, that is, the use of media as resources or didactic auxiliaries for the betterment of the teaching-learning processes. In media education, the task of critical training and knowledge of reality is conducted –the media are the objects of study-, although both disciplines can be perfectly compatible and even, as we will show in a moment (Gómez Galán, [15] [16]), complementary.

Currently, media education is essential. To educate in order to form a society that proves to be critical when facing the power of techno-media messages –in all its current characteristics, in the 21st century– is one of the essential tasks in any educational context. Teaching professionals need solid training in this discipline, and there is nothing better for this than basing themselves on the knowledge that involves the same constituent pillars, that rose during the whole of the 20th century. The milestones reached throughout this ample space of time –and above all from the 1920s to the 1980s– are perfectly valid currently, and what's more, they are more important than ever for any educator since the media has multiplied its power and social presence thanks to the techno-media convergence.

The epistemological structure of media pedagogy or education should not be very different currently, in the society of the digital revolution, from what it was in its origins, since the objectives are similar. Above all we find ourselves with the need for educational processes that facilitate socialization and democratic growth of countries, the main objective of this specialty of the education sciences. We can talk about the updating of methods and didactic models precisely for the techno-media explosion in which we find ourselves, but from a theoretical base the fundamentals are identical. Of course it makes dialogue even more necessary with other sciences or fields of knowledge, even in the realm of pedagogy –let's think about, for example, the necessity of establishing more and more open channels with educational technology-- the most important of all is that the collection of what we have obtained until now continues being perfectly valid.

## 3. The Processes of Human Communication and the Digital Revolution: Analysis from a Cultural and Educational Perspective

The digital revolution is a consequence of the established efforts by human beings for improving communication processes. Currently we find ourselves inside the historical process of the evolution and development of the means of communication, that our species, since our origins, has used to transmit all types of information. It's true that there have been decisive moments -and possibly we find ourselves in one of them- but the process has been linear, consistent with the technological and cultural advances of humanity. It is clear that in moments of revolution we can identify causes and consequences of the historic transformation and in this way better explain the phenomena. And the issue that we are facing now is no exception.

The emergence of means of communication took place in the industrial revolution, and it came alongside vast progress in many fields and dimensions –social, economic, political, cultural, technological, etc.-, they have been culminated in recent decades in that which we have denominated, as we already explained, technology-media convergence, which happened thanks to the digital revolution –let's place this within the industrial revolution, like Castells [17] would have defended in his time, or, what only the future will determine, signifies the start of a new revolution whose consequences we are even further from glimpsing- (Gómez Galán [18]). But, certainly, human communication structures continue on in a very similar way.

There is no doubt that throughout history, fundamental moments in the development of the means of human communication have been made, like what the birth of writing could have been –precisely the cusp of the neolithic revolution, which, in its origins supposed, naturally, a radical transformation of the means of production– or that of the printing press, on the verge of the industrial revolution. Throughout hundreds of years, a gradual growth of the possibilities of communication happened, and the characteristics and means of doing so were diversified. In the $20^{th}$ century, accompanied by the technological development that allowed this process, we find the existence of multiple communicative means that employ the most diverse systems and languages: writing, sound, images, multimedia techniques, etc. And precisely at the end of the last century, however, thanks to the development of information and telematic technology, technologies centered on the automatic treatment of information and the transmission of it through networks –whose initial objective was the optimization of management--, another decisive moment in the transformation of means of communication was being forged: the digitalization of information, which brings media to a convergence. Today, in the first half of the $21^{st}$ century, we are witnessing, like astonished spectators, a new revolution of the human communicative processes.

The fundamental difference that we can find is that the process of digitalization has enormously strengthened interactivity with and among media. It has given place to that which is called the sender-receiver model. In many cases, the division between who sends and who receives the information and who acquires it is not clear. Today, users appear as active participants of the product, but also as creators and not just consumers. As such, and as we have pointed out, although the main channels of communication continue being controlled by powerful media groups, the recipient of the messages is offered the illusion of believing him or herself to be the active participant in the process. Naturally he or she only has the possibility of participating in messages of low intensity, imperceptible in the control of the media, but sufficient enough so that he believes himself to be important and decisive within the dynamic.

In the face of this new reality, we believe that an evolution of the methods and didactic models is necessary, breaking from the traditional bases, which support it, that we will present later on. One of the strategies that we consider most adequate, that would be only an example, is precisely the dialogue within educational technology, about combining within the processes of teaching-learning the presence of the media with didactic resources—all citizens, regardless of age, are users and consumers of them—as an object of study. The facilities that the digital paradigm offers us for the use of the media in the classroom grant us a decisive advantage in order to strengthen education in it.

Let's look at an example. Let's think, without going any further, about the possibilities of the press. It would not be the same to have a newspaper available in the classroom with which to develop a concrete didactic activity, as accessing editions on the Internet, which are permanently updated, of four or five newspapers, to take advantage of all the possibilities that they give us, in order to, let's state our case, do a critical analysis of the news (reading the same news in different papers, measuring the space dedicated by each of them, the treatment, utilized sources, and contrast among them, etc.) The Internet, it's clear, would have made this task easier (and in the opposite way we would have needed to acquire multiple issues, to highlight, to trim the content, etc.) but, in essence, the process of study and pedagogical use of ICT –today turned into powerful media—will be able to support itself on the classical structure of media education.

We can make all this extend to any computer tool or new telematic tool, without going any further than social networks, like Facebook or Twitter, which are currently so popular. We could utilize any of them for didactic development with the goal of strengthening interactivity, group work, creativity, motivation, etc., and surely the results—if everything has been adequately designed based on pedagogy—will be positive. However, and as we widely defended on another occasion (Gómez Galán, [19]), we cannot stop the margin of impact that these have on society, and especially their influence on childhood and youth. We should take advantage of their use as didactic resources in order to turn them into parallel objects of study, of analysis of their characteristics, of the use that the student makes of them in his life, of the advantages, but also possible dangers that they contain. It would even be adequate to first employ social networks under educational control as preparation for the knowledge and management of the most socially extended, of which, without a doubt, children and young adults will turn into, sooner or later, and then into consumers. We would talk about, therefore, an integral education in which ICT/media is contemplated in their authentic social and educational dimension, in the sphere of techno-media convergence.

## 4. Theoretical and Practical Bases of the Pedagogical Paradigm of Media Education

What can, therefore, the pedagogical paradigm of media education, with the goal of completing adequate digital literacy education, provide us? Without a doubt, some solid theoretical and practical structure, gotten thanks to multiple educational experiences which happened for almost a century and in many countries — although it fundamentally developed in the Anglo-Saxon sphere—, which we consider most important and decisive in the authentic education for this extremely important social dimension. Because, to educate about the media is to educate about digital literacy, definitely. The separation among media—alluding to the traditional mass media communication means— and ICT —as technologies of management of information and communication— already does not exist in the context of techno-media convergence. If we can update the methods and models, as we have seen in the practical example presented in the previous point, the objectives of media education today are similar to those pursued in previous decades of the digital revolution. It is to educate in the realm of communication.

We will focus, therefore, on establishing what the educational bases for the media are, which is invaluable for a correct digital literacy education. The first one we will have if we were to do a global review of the evolution of this discipline, is from the first decades of the 20$^{th}$ century until approximately the current era. As it is not the objective of this article to make a detailed description of this process—as the most simple description can be found by consulting Rivoltella [20], although we particularly defend an evolution with distinct phases—we will summarize by saying that, throughout this wide space of time there were various objectives to pursue: (1) To describe the educational potential of each means of communication; (2) to study the media in a thematic and methodological multi-disciplinary scope (from sociology, psychology, anthropology, etc., through the use of discourse analysis, image analysis, comparative studies, etc.); and (3) to complete a critical lecture on the techno-media products, naturally, from a pedagogical and educational perspective, as we have presented.

The first of these objectives, centered on the didactic scope, which is fundamental in the whole process of digital literacy education, supposes something that no educational professional should ever forget: the important thing is not the media, but how it is used, the characteristics of the student body, and other related variables. The majority of students developed during so many decades demonstrated that an in-undeniable didactic value exists in each medium, but always some determined conditions for their correct usages are formed. The educational value of a medium is always relative, it can be extremely beneficial but at the same time, if the usage and the context are not favorable for it, it can even be counterproductive. And the long trajectory of the studies centered on this objective, from these dates until today, have not made but corroborated these initial results. Today, in the context of techno-media convergence, this choice multiplies exponentially, and should always be kept in mind by any educator.

The second objective, the dialogue with other disciplines and the use of multiple methodologies, is the path for the advance of an authentic media education, in the complexity that its objective of study covers. There have existed, throughout this whole extensive space of time, multiple studies and theories which support and intercommunicate with each other. As some authors have defended (Wartella y Reeves [21]; Mariet & Materman [22]; Gonnet [23]; and Rivoltella [24]), in a review that they did on the characteristics of media education in distinct historical eras, the theoretical configuration of this discipline is found horseback riding between the educational sciences and communicate on sciences, participating in the same way in distinct arenas like semiotics, sociology, cultural studies, linguistics, etc. It is therefore about the specialty of boundaries with multiple methodological possibilities within the sphere of research. And this, for the objective of the present article, is summarily important because it allows us to defend the theoretical review completed throughout different eras, which is extremely valuable currently, in order to define the needs of digital literacy education with a solid background.

In this way, media education has been supported —as Rivoltella defends [25] or we have studied ourselves (Gómez Galán, [26])— by a great convergence of studies like those of Hovland [27], Gerbner ([28] [29]), Gerbner, Gross, Morgan & Signorielli [30] or Noelle-Neumann ([31] [32]), in the realm of communication theory; McLuhan [33], Hall & Whannel [34], Herman & Chomsky [35], Giroux [36] or Giroux & Shannon [37], in cultural studies; Baumann ([38] [39] [40]), in sociology; Eco [41], in semiotics; Dewey [42], Freire ([43] [44]) or Freinet [45], in pedagogy itself; etc. And this would only be an example of an extremely wide panorama.

This has given a place for multiple dimensions within media education, and the ways of understanding its functions. Although everyone participating, naturally, in a general and common vision of creating critical attitudes within the student body in the face of techno-media power, which would give a response to the third objective presented. It would bring us enough time and space to describe, not just superficially, that which we will indicate that, in our judgment, the main contributions have been those completed directly by media education specialists who have collected the essentials from the previous studies in order to create a theoretical structure that is rather defined and specific for the discipline. We consider that the research of Masterman ([46] [47] [48]), Buckingham ([49] [50]), Alvarado & Boyd-Barrett [51], Aufderheide & Firestone [52], Kellner ([53] [54]), Silverblatt [55], Potter ([56] [57]), Hart [58], Messaris [59], or Ferguson ([60] [61]), for example, makes a paradigmatic example of it. The tasks completed by these authors synthesize the best conclusions obtained by the majority of specialists that, in one way or another, have been close to the issue. It's about, therefore, authentic theoretical constructs of the discipline, from those that we should start from, without a doubt, in order to develop a correct digital literacy education.

## 5. Principles of Digital Literacy Education in Techno-Media Convergence

Keeping in mind, therefore, the nucleus of media education, the fruit of our research, we are in a condition to present the fundamental principles of every current process of digital literacy education, and what we consider invaluable currently for the study of education in society of the digital revolution and techno-media convergence, and an authentic education for the citizens of the 21$^{st}$ century.

The first response starts with the same question, of why we educate for technology and means of communication today, and how to do it by overcoming the failures that, until now, the aforementioned company has supposed. Masterman [62], from a traditional media education standpoint, culminated this, which is perfectly valid currently, considered that it is invaluable for: (1) The elevated rate of consumption of media and its saturation in current society; (2) the ideological importance of them and their influence as actions of awakening; (3) the increase of the manipulation and production of information and its propagation through them; (4) the progressive entrance of the media in basic democratic processes; (5) the prevailing protagonism of communication by means of the image; (6) the importance of educating students so that they can handle the requirements of the future and (7) the fast increase of national and international pressures with the end of privatizing information.

All of these are not just highly topical reasons, but also are strengthened profoundly by the power today of communicative processes. Although it seems that we find ourselves 25 years after these conclusions, in a distinct techno-media panorama, in the background—and how we have previously defended it—the situation is similar. We would say that it is even more urgent. We would contemplate, for example, the capacity of the current media to manipulate information and to create *virtual realities* that are so difficult to distinguish from true reality.

Once this first question is answered, we would affront the second, which would be focused on determining which benefits we can reach with a digital literacy education originated on the base of media education, further off in the future than the current extended instrumental and technological base. The studies completed for decades, focused on this paradigm, have presented distinct possibilities, but fortunately there is a certain consensus on the advantages that it can report to us. We will select a few of them –conveniently adapted from and contemplated with their own criteria– that have been accepted by the majority of specialists unanimously and are found defended, in large part, by Masterman [63]: (1) To educate for the media strengthens the democratic structures of society; (2) media education allows for critical education in people, within a process that is produced during one's entire life; (3) it causes changes and innovation in education, since it analyzes a reality in continual transformation; (4) media education is a collaborative process, that requires the full intervention of the whole educational community which lays down its bases; (5) the development of the critical capacity of students is its first objective, which implies the overcoming of memorization learning; and (6) media education supposes the application of new pedagogical paradigms:

formative education, student-teacher relationships based on the dialogue, didactic methodologies that spur the participation and activity of students, etc.

Basically, we can speak about the settlement of the most open and democratic pedagogies, that promote responsibility and self-learning of students. These last two points already suggested a new didactic for its teaching, and although today there is not a defined methodological framework –perhaps because this should be a function of the environment and the socio-cultural characteristics of the educational context where it is developed, which makes homogeneity impossible– we could point out some considerations for the educational practice with technology and communication media: (1) It is essential that they are considered as systems of symbols or signals, not like reality itself –from the perspective they should be studied–; (2) it's necessary to develop in the student the use of analytic instruments that allow him to establish his own criteria of study; and (3) it should be an education based on research, not on imposition.

We finish by offering, therefore, the main principles reached by *media education*, in its many years of application and development, with perfect validity in current society, in the realm of techno-media convergence and digital literacy education. Many recent studies continue insisting on the importance of this discipline, like that of Balazak [64], Faiola, Davis, y Edwards [65], Kratka [66], King [67], Amat [68], Petranova [69] or Zavala [70], which show the interest that continues awakening the scientific community. If we use these principles to form the new generations in the context of a hyper-communicated world and dominated by the flows of information, with all that it implies and that we have described, we will be contributing without a doubt to the progress of our society.

## 6. Discussion and Conclusions

To achieve digital literacy nowadays a media education is required. Independently of the perspective from which the study and analysis is undertaken, especially if from a multidisciplinary perspective, it is clear that we are passing through a new stage in the occidental civilization. After the neolithic and then the industrial revolution, the digital revolution is bringing about the third great stage: the information society and/or the communication society. This is leading to the appearance of a new social, political, cultural and economic order whose evolution and results are as yet unpredictable and uncertain.

The digital revolution breaks the barriers of space and time, and information has become the most valuable commodity. The totality of the ICT, which in its day had very specific aims in the military, economic and scientific areas has transformed into consumer and social ones. The powerful mass media, that little by little increased its presence throughout the 20th century, took advantage of this situation to increase, in an unheard of way, its power and domination, with all that that implies at the social level. The new digital reality, championed by the network of networks –i.e.: Internet– favours the convergence of all communicative media in existence and becomes the most powerful instrument, resource, means and message in globalized society. An instrument, let us not forget, in the hands of its creators and masters, who have invested so much in its birth and development.

Therefore, far from having communicative freedom and independence, the techno-media society is dominated by economic, political and social elites –ever closer to an evident and common interest; influence over and dominance of the citizens–. They are the owners of the system, the creators of technology, the masters of information. However, this new Orwellian big brother is unnoticed in its watching, it is much more subtle, complex, expansive. We have the appearance of complete liberty in social networks, in the transfer of information, in interpersonal communication, etc, but, nevertheless, it is controlling everything. This is natural because far from being a system which is altruistic and concerned about the total development of our civilization, we are in its eyes, above all, consumers, voters and numbers.

In this new reality –the same reality as always but neatly disguised– education could be the catalyst of change, not only continuity. If the educational structures are serving the ICT then they are serving its masters. Therefore, it is imperative to break this tendency and analyze it in a real and critical way, in all its complexity and dimensions. Only through education will it be possible for a better world, the growth of values and solidarity in our society, to fight against so much injustice which exists on our planet and to seek, specifically, the preservation of it, against the threat of an economy which seeks huge and easy benefits for the elite. In school its origin, means and development must be analyzed and citizens prepared to confront the power and influence of the techno-media messages. This is not being done, and we have numerous doubts that this will ever be done. We have to consider, first of all, if current institutions are independent enough to carry this out, but this would be another study beyond the space of this article.

Let us not be deceived, it is difficult to find new problems. The current ones originated, directly or indirectly, in the historical processes from which they surged. In history it is impossible to speak about novelties which are radically independent from what has come before. Taking this premise as our starting point, and as we have argued, we can observe in the educational area, media education not only belongs to a past which can seem far away, but it is concerned with an educative model of complete and urgent importance. Naturally, a continuous development of this model adapted to changing social characteristics, with its base and foundation, gestated throughout the 20th century, is perfectly valid nowadays. This is something which should be essential in the formation of teaching professionals in search of an authentic education for the needs of the contemporary world, and a fundamental pillar of a correct literacy of a citizen of the 21st century. It reaffirms the value which education has for the media as one of the main instruments which will allow us to transform and improve current society, and undertake from grass roots effective digital literacy.